\documentclass[%
 reprint,
superscriptaddress,
 amsmath,amssymb,
 aps,
pra,
nofootinbib,
noeprint
]{revtex4-2}

\usepackage{graphicx}
\usepackage{bm}
\usepackage{amsmath,amssymb}
\usepackage{braket}
\usepackage{breqn}
\usepackage{xcolor}
\usepackage{float}
\usepackage{multirow}
\usepackage{mathtools}
\usepackage{color}

\begin{document}
\title{Optical cycling in polyatomic molecules with complex hyperfine structure}
\author{Yi Zeng}
\email{yzeng@caltech.edu}
\affiliation{California Institute of Technology$,$ Division of Physics$,$ Mathematics$,$ and Astronomy$,$ Pasadena$,$ CA 91125$,$ USA}

\author{Arian Jadbabaie}
\affiliation{California Institute of Technology$,$ Division of Physics$,$ Mathematics$,$ and Astronomy$,$ Pasadena$,$ CA 91125$,$ USA}

\author{Ashay N. Patel}
\affiliation{California Institute of Technology$,$ Division of Physics$,$ Mathematics$,$ and Astronomy$,$ Pasadena$,$ CA 91125$,$ USA}

\author{Phelan Yu}
\affiliation{California Institute of Technology$,$ Division of Physics$,$ Mathematics$,$ and Astronomy$,$ Pasadena$,$ CA 91125$,$ USA}

\author{Timothy C. Steimle}
\affiliation{California Institute of Technology$,$ Division of Physics$,$ Mathematics$,$ and Astronomy$,$ Pasadena$,$ CA 91125$,$ USA}
\affiliation{School of Molecular Science$,$
Arizona State University$,$
Tempe$,$ Arizona 85287$,$ USA}

\author{Nicholas R. Hutzler}
\affiliation{California Institute of Technology$,$ Division of Physics$,$ Mathematics$,$ and Astronomy$,$ Pasadena$,$ CA 91125$,$ USA}

\date{\today}
\begin{abstract}
We have developed and demonstrated a scheme to achieve rotationally-closed photon cycling in polyatomic molecules with complex hyperfine structure and sensitivity to hadronic symmetry violation, specifically $^{171}$YbOH and $^{173}$YbOH. We calculate rotational branching ratios for spontaneous decay and identify repumping schemes which use electro-optical modulators (EOMs) to address the hyperfine structure. We demonstrate our scheme by cycling photons in a molecular beam and verify that we have achieved rotationally-closed cycling by measuring optical pumping into unaddressed vibrational states. Our work makes progress along the path toward utilizing photon cycling for state preparation, readout, and laser cooling in precision measurements of polyatomic molecules with complex hyperfine structure. 
\end{abstract}

\maketitle

\subsection{Introduction}
 
Recent advances in the cooling and trapping of increasingly complex molecules \cite{Fitch2021Review,Augenbraun2023PolyLCReview} are enabling a multitude of applications that leverage molecular complexity for applications in quantum science and precision measurement \cite{Carr2009,Bohn2017,DeMille2017,Safronova2018,Hutzler2020Review}. One avenue for increased complexity is the use of molecules with a heavy nucleus and non-zero nuclear spin, $I$. Precision measurements with these molecules can probe new physics related to the nucleus, such as charge-parity (CP) violating nuclear moments \cite{Safronova2018}, nuclear-spin-dependent parity violation (NSD-PV) \cite{Wood1997,Altunta2018NSD}, and measurements of nuclear structure \cite{Udrescu2021RaF,RadMol2023}.

    One example is the odd isotopologues of the linear polyatomic molecule YbOH, which are promising candidates for probing symmetry-violating physics in the hadronic sector: $\prescript{171}{}{}$YbOH for parity violation~\cite{Norrgard2019}, and $\prescript{173}{}{}$YbOH for the CP-violating nuclear magnetic quadrupole moment (NMQM)~\cite{Flambaum2014,Kozyryev2017PolyEDM}. Interactions of core-penetrating  valence electrons with the heavy Yb nucleus enhance sensitivity to symmetry violation~\cite{Maison2019,Denis2020}, and the quadrupole shape deformation of the Yb nucleus provides further collective enhancement of CP-violating moments~\cite{Flambaum2014}. Additionally, the vibrational bending mode of YbOH exhibits closely spaced, opposite parity levels. These parity doublets allow for control of molecular polarization in modest electric fields~\cite{Jadbabaie2023YbOHBend}, suppression of systematic errors in precision measurements~\cite{Kozyryev2017PolyEDM,Anderegg2023CaOHSpin}, and can be tuned even closer to degeneracy by modest magnetic fields, enhancing sensitivity to NSD-PV~\cite{Norrgard2019}. Bending modes are one example of parity doublet structures that are generic to polyatomic molecules, and which exist independently of the electronic structure, thereby enabling their combination with efficient optical cycling and laser cooling ~\cite{Kozyryev2017PolyEDM,Mengesha2020YbOHBranching,Augenbraun2020YbOHLC,Zhang2021BR,Hutzler2020Review}.

The additional complexity of polyatomic molecules presents both opportunities and challenges. In particular, the increased number of vibrationally excited states in the electronic ground state provide many pathways for spontaneous decay after optical excitation. Molecules that decay to ``dark states'' not addressed by repumping lasers are lost from the optical cycle~\cite{DiRosa2004,Fitch2021Review}. Addressing rotational and vibrational branching to dark states has enabled laser cooling and trapping of both diatomic and polyatomic molecules~\cite{Fitch2021Review,vilas2022magneto}.  However, eliminating dark states in species with large and complicated hyperfine structure, which is a byproduct of many molecules with spin on a heavy nucleus,  poses additional challenges~\cite{Kogel2021BaFBR}. In particular, hyperfine splittings will have a multiplicative effect on the number of existing rotational and vibrational dark states, with frequency splittings large enough such that bridging with modulators becomes a challenge.

In this manuscript, we report the design and experimental realization of a scheme for achieving rotationally closed cycling of $\prescript{171}{}{}$YbOH ($I_{\textrm{Yb}}=1/2$) and $\prescript{173}{}{}$YbOH ($I_{\textrm{Yb}}=5/2$). With only two modulation frequencies on a single laser, we achieve photon cycling with closure of rotational, spin-rotational, and hyperfine structure. Furthermore, such a scheme should be broadly applicable to other molecules with similar structure, especially with the implementation of additional techniques, such as computer-generated holography~\cite{Holland2021holo}.

\subsection{Branching ratio calculation}

In this manuscript, we address the branching of spontaneous decays $\Tilde{A}^2\Pi_{1/2}(0,0,0)\rightsquigarrow \Tilde{X}^2\Sigma^+(0,0,0)$.  Here $\Tilde{A}^2\Pi_{1/2}(0,0,0)$ is the vibrational ground state of the excited electronic state $\Tilde{A}^2\Pi_{1/2}$, and $\Tilde{X}^2\Sigma^+(0,0,0)$ is the vibrational ground state of the ground electronic state $\Tilde{X}^2\Sigma$.  We will often refer to the states as simply ``$A$'' and ``$X$,'' respectively, and if the vibrational quantum numbers $(\nu_{\mathrm{Yb-O\, stretch}},\nu_{\mathrm{bend}},\nu_{\mathrm{O-H \, stretch}})$ are omitted, it is assumed that we mean the ground vibrational state $(0,0,0)$. The vibrational branching ratio of $A\rightsquigarrow X$ decay has been previously measured~\cite{Zhang2021BR} to be 89.44(61)\%.  Here, we are primarily concerned with the rotational, spin-rotational, and hyperfine branching, which are significantly more complicated in laser-coolable species with non-zero nuclear spins on the metal and the ligand~\cite{Kogel2021BaFBR}.

The branching ratios are calculated using effective Hamiltonians \cite{Brown2003}. The energies and eigenvectors within the $X$ and $A$ states are obtained by diagonalizing Hamiltonian matrices in the Hund's case (a) basis, and then transition dipole moments (TDMs) are calculated between the eigenvectors of the two states. For simplicity, the hydrogen nuclear spin from -OH was not included in the basis set, as the hyperfine splitting is not optically resolved \cite{nakhate2019pure}.  However, we must remember that it is present when calculating level degeneracies.

The effective Hamiltonians used for the two states are~\cite{Pilgram2021YbOHOdd}:
\begin{dmath}
H^{\mathrm{eff}}_X = B\bm{N}^2 + \gamma \bm{N} \cdot \bm{S}+ b_F \bm{I} \cdot \bm{S} + c(I_z S_z - \frac{1}{3} \bm{I} \cdot \bm{S}) + e^2Qq_0 \frac{3I^2_z-\bm{I}^2}{4\bm{I}(2\bm{I}-1)},
\end{dmath}
\begin{dmath}
H^{\mathrm{eff}}_A = A L_z S_z +B\bm{N}^2 + \frac{1}{2}(p+2q)(J_- S_- e^{+2i\phi}+J_+S_+e^{-2i\phi}) + h_{1/2} I_zL_z
-\frac{1}{2} d(S_+I_+e^{-2i\phi}+S_-I_-e^{+2i\phi})+  e^2Qq_0 \frac{3I^2_z-\bm{I}^2}{4\bm{I}(2\bm{I}-1)}.
\end{dmath}
Here, $\bm{S}$ is the Yb-centered electron spin, $\bm{I}$ is the Yb nuclear spin, and $\bm{N}$ is the total non-spin angular momentum. All angular momentum subscripts ($z, \pm$) denote molecule frame components. Spin-orbit $A$, rotation $B$, spin-rotation $\gamma$, and $\Lambda$-doubling $(p+2q)$ are present in all isotopologues. Both $\prescript{171}{}{}$Yb($I\,{=}\,1/2$) and $\prescript{173}{}{}$Yb($I\,{=}\,5/2$) have nuclear spins, which give rise to additional hyperfine parameters, namely orbital hyperfine $a$, Fermi contact $b_F$, spin-dipolar $c$, and parity-dependent dipolar $d$. In the excited state, the diagonal hyperfine shifts are determined by an effective parameter $h_{1/2}$, which can be written as $h_{1/2} = a - (\frac{b_F}{2} + \frac{c}{3})$~\cite{Pilgram2021YbOHOdd}. Further, $\prescript{173}{}{}$Yb also has an electric quadrupole moment, which gives rise to the term $e^2Qq_0$. The exact parameter values used are taken from ref.~\cite{Pilgram2021YbOHOdd} and listed in table~\ref{paratab}. 

After diagonalizing the Hamiltonians, the eignenvectors are labeled according to the Hund's cases that best represent their structure: Hund's case ($b_{\beta S}$) for $X$, and Hund's case ($a_{\beta J}$) for $A$ \cite{Pilgram2021YbOHOdd}. The labels are simplified to include only the quantum numbers relevant to this manuscript: $\ket{NGF}$ for the $X$ state and $\ket{J(P)F}$ for the $A$ state. Here, for the $X$ state $\bm{G} \,{=}\, \bm{I}+\bm{S}$ results from the hyperfine interaction between $\bm{I}$ and $\bm{S}$, while for the $A$ state $\bm{J}$ results from both molecule rotation and spin-orbit coupling and $(P)$ is the parity label for the parity doublets resulting from $\Lambda$-doubling. The total angular momentum $\bm{F}$ is given by $\bm{F} \, {=} \, \bm{G}+\bm{N}$ in the $X$ state and $\bm{F} \, {=} \, \bm{J}+\bm{I}$ in the $A$  state. We will use double primes to denote ground state quantum numbers (e.g. $N''$) and single primes to denote excited state quantum numbers (e.g. $J'$).

TDMs are calculated between the eigenvectors by representing all states in the Hund's case (a) basis and performing computations only in this basis. Case (b) quantum numbers labels $N'', G''$ are assigned to the ground state eigenvectors by identifying patterns in the eigenvalues. To obtain branching ratios and account for the degeneracy of $M_F$ sublevels, the TDMs are summed over light polarization $p$ and ground state $M_F''$ levels, and averaged over excited state $M_F'$ levels. We obtain:
\begin{dmath}
\frac{1}{2F'+1} \sum_{p,M_F'',M_F'} \bra{N''G''F''}T^1_p(d)\ket{J'(P)F'}.
\end{dmath}
This way, the branching ratio derived by squaring the TDM is naturally normalized, which means that the branching ratios originating from the same excited state add up to 1. Note, we use the same reduced matrix element convention as ref.~\cite{Brown2003}.

\begin{table}[h]
\begin{ruledtabular}
\begin{tabular}{llll}
 State                               & Parameter      & $^{171}$YbOH  & $^{173}$YbOH  \\ \hline
\multirow{5}{*}{$\Tilde{X}^2\Sigma^+(0,0,0)$}  & $B$         & \phantom{$-$}0.245    & \phantom{$-$}0.245    \\ 
                                & $\gamma$  & $-$0.00270 & $-$0.00270 \\ 
                                & $b_F$      & \phantom{$-$}0.228    & $-$0.0628  \\ 
                                & $c$         & \phantom{$-$}0.0078   & $-$0.00273 \\ 
                                & $e^2Qq_0$ & N/A      & $-$0.111   \\ \hline
\multirow{6}{*}{$\Tilde{A}^2\Pi_{1/2}(0,0,0)$} & $A$         & \hspace{-2.5mm}1350     & \hspace{-2.5mm}1350     \\ 
                                & $B$         & \phantom{$-$}0.253    & \phantom{$-$}0.253    \\ 
                                & $p+2q$      & $-$0.439   & $-$0.438   \\ 
                                & $h_{1/2}$         & \phantom{$-$}0.0148   & $-$0.00422 \\ 
                                & $d$         & \phantom{$-$}0.0320   & $-$0.00873 \\ 
                                & $e^2Qq_0$ & N/A      & $-$0.0642  \\ 
\end{tabular}
\caption{Relevant parameters, from \cite{Pilgram2021YbOHOdd}, in wavenumbers ($\text{cm}^{-1}$) for the $\Tilde{X}^2\Sigma^+(0,0,0)$ and $\Tilde{A}^2\Pi_{1/2}(0,0,0)$ states of $\prescript{171}{}{}$YbOH and $\prescript{173}{}{}$YbOH. }
\label{paratab}
\end{ruledtabular}
\end{table}

\begin{figure}[h]
\centering
    \includegraphics[width=\columnwidth]{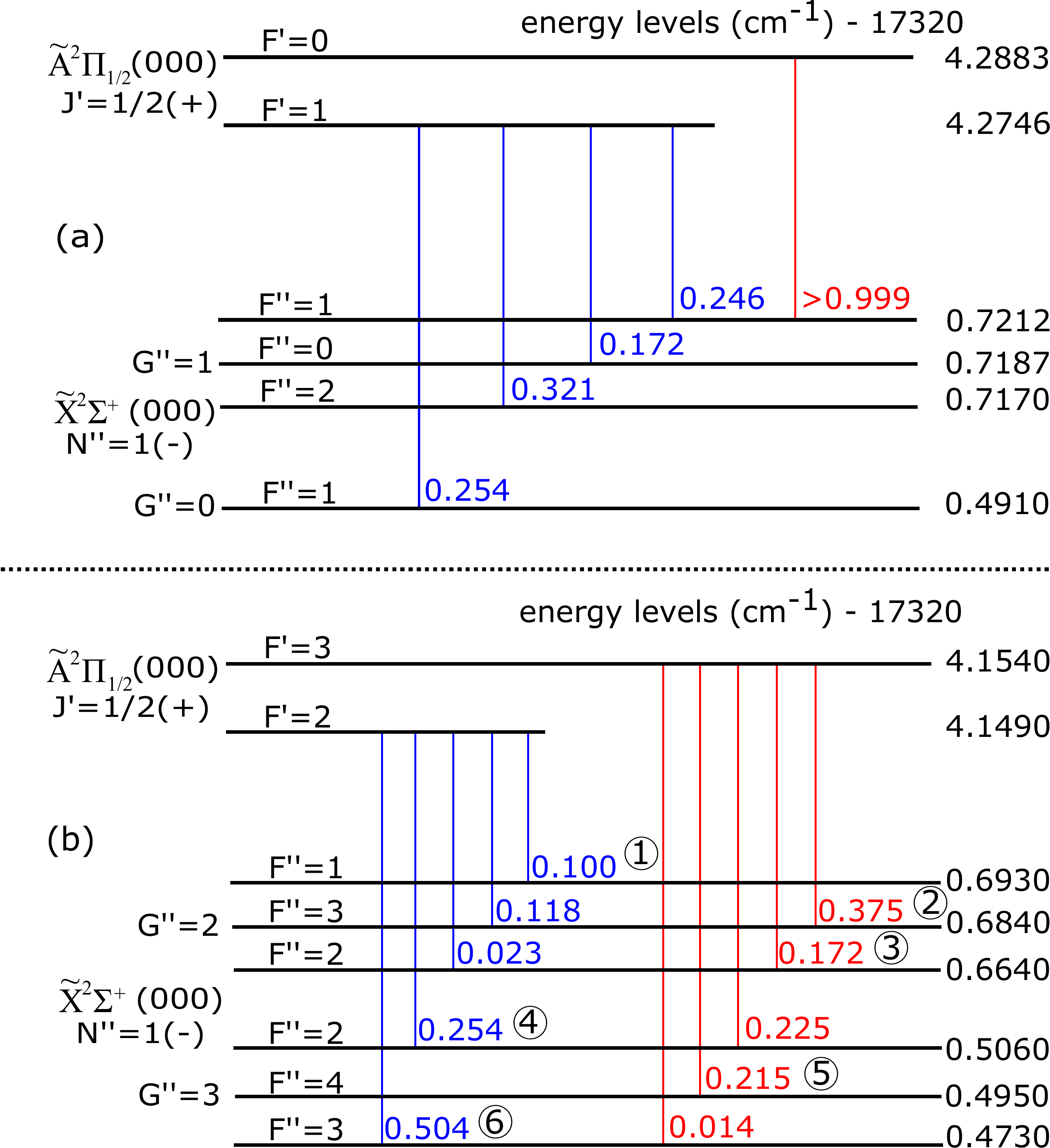}
    \caption{Calculated rotational branching ratios for (a) $\prescript{171}{}{}$YbOH and (b) $\prescript{173}{}{}$YbOH. Numbers 1-6 in circles label the transitions targeted for achieving rotationally closed cycling.}
    \label{brfig}
\end{figure}

The calculated branching ratios are shown in fig.~\ref{brfig}. Note that these branching ratio numbers add up to 1 only within the same vibration level, and they will be referred to as rotational branching ratios. The total branching ratio of a transition can be derived by multiplying the rotational branching ratio by the vibrational branching ratio, which is known for $\prescript{174}{}{}$YbOH \cite{Zhang2021BR}, and expected to be the same for $\prescript{171,173}{}{}$YbOH.

\subsection{Method for achieving rotationally closed cycling}

We achieve rotationally-closed optical cycling via $N''\,{=}\,1\rightarrow N'\,{=}\,0$ type transitions~\cite{stuhl2008magneto}, specifically $N''\,{=}\,1\rightarrow J'\,{=}\,1/2(+)$ for YbOH (since $N'$ is not well-defined in $A$, $J'=1/2$ corresponds to the lowest rotational level in $A$). Based on our calculations, rotationally closed cycling is straightforward to achieve for $\prescript{171}{}{}$YbOH because there is a single transition, $F''\,{=}\,1\rightarrow F'\,{=}\,0$, with calculated rotational branching ratio of $>0.999$. The excited hyperfine splitting of $\sim$400~MHz is sufficiently large, compared to the experimental linewidths of $\sim$50~MHz , such that off-resonant excitation to $F'\,{=}\,1$ can be avoided, which is not the case in $\prescript{173}{}{}$YbOH.

\begin{figure*}[ht]
\centering
\includegraphics[width=0.65\textwidth]{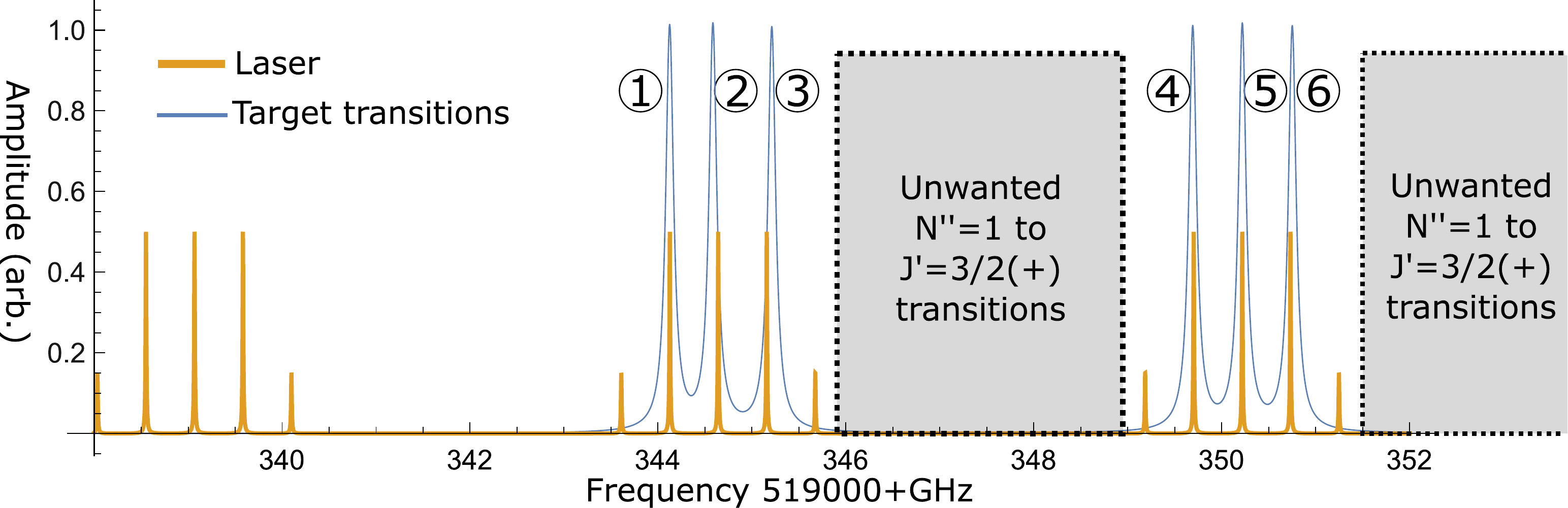}
\caption{Scheme for EOMs to address ${}^{173}$YbOH branching. Blue (thin) lines are the target transitions we want to address. From left to right they correspond to transitions in fig.~\ref{brfig}(b) labeled 1-6.  Yellow (thick) lines are generated by the pump laser and EOMs, which reflects the frequency spacings and relative amplitudes of the sidebands generated EOMs as verified by using a Fabry-Perot cavity. As shown, the scheme used can address all the target transitions within their linewidths while avoiding the unwanted lines that can cause leakage to dark states.}
\label{eofig}
\end{figure*}

In order to fully address rotational branching for $\prescript{173}{}{}$YbOH, we must address all 6 ground hyperfine levels, as the excited $F'\,{=}\,2$ and $F'\,{=}\,3$ states are separated by $\sim$150 MHz, which can result in off-resonant excitation when slightly power-broadened. We address the ground levels by generating sidebands on a single laser using two fiber electro-optical modulators (EOM) used in series\footnote{EOSPACE PM-0S5-10-PFA-PFA-1154-UL-SOP125mW}. Fortunately, the transitions are spaced such that it is possible to use only two EOMs, each with a single sinusoidal drive, to address all of transitions within the slightly power-broadened linewidth, as shown in fig.~\ref{eofig}.

The two EOMs are used sequentially with input from an 1154~nm seed laser, and the output is then amplified and doubled to produce the needed visible light at 577~nm. Extra, unwanted sidebands are produced from both the EOMs and the doubling crystal, and the latter also modifies the sidebands due to the non-linear nature of second harmonic generation. These sidebands can accidentally drive undesirable transitions that cause leakage by optically pumping $N''\,{=}\,1$ population into rotational dark states within the same vibrational manifold.

Here, we were able to avoid leakage-inducing transitions by using modulation frequencies set such that the three intense, unused sidebands, as well as smaller, higher-order sidebands, are sufficiently far away from unwanted transitions, as shown in fig.~\ref{eofig}. Note, however, that the complex and congested spectrum \cite{Pilgram2021YbOHOdd} of these molecules means that off-resonant excitation of unwanted lines cannot be ignored, and our method takes them into account when calculating the number of scattered photons.

\subsection{Measuring the number of scattered photons}

We experimentally study  optical cycling using a cryogenic buffer gas beam (CBGB)~\cite{Hutzler2012CBGB} of YbOH. We create cold, slow beams of YbOH using methods similar to those described elsewhere~\cite{Jadbabaie2020,Pilgram2021YbOHOdd,Jadbabaie2023YbOHBend}.  We use a $\sim$1.5 K cell with 2 SCCM flow of helium buffer gas, and enhance the production of YbOH using optically-driven chemical reactions~\cite{Jadbabaie2020}.  We use CW lasers to scatter photons, and the resulting laser-induced fluorescence is monitored with photo-multiplier tubes (PMTs).

\begin{figure*}[ht]
\centering
\includegraphics[width=1.2\columnwidth]{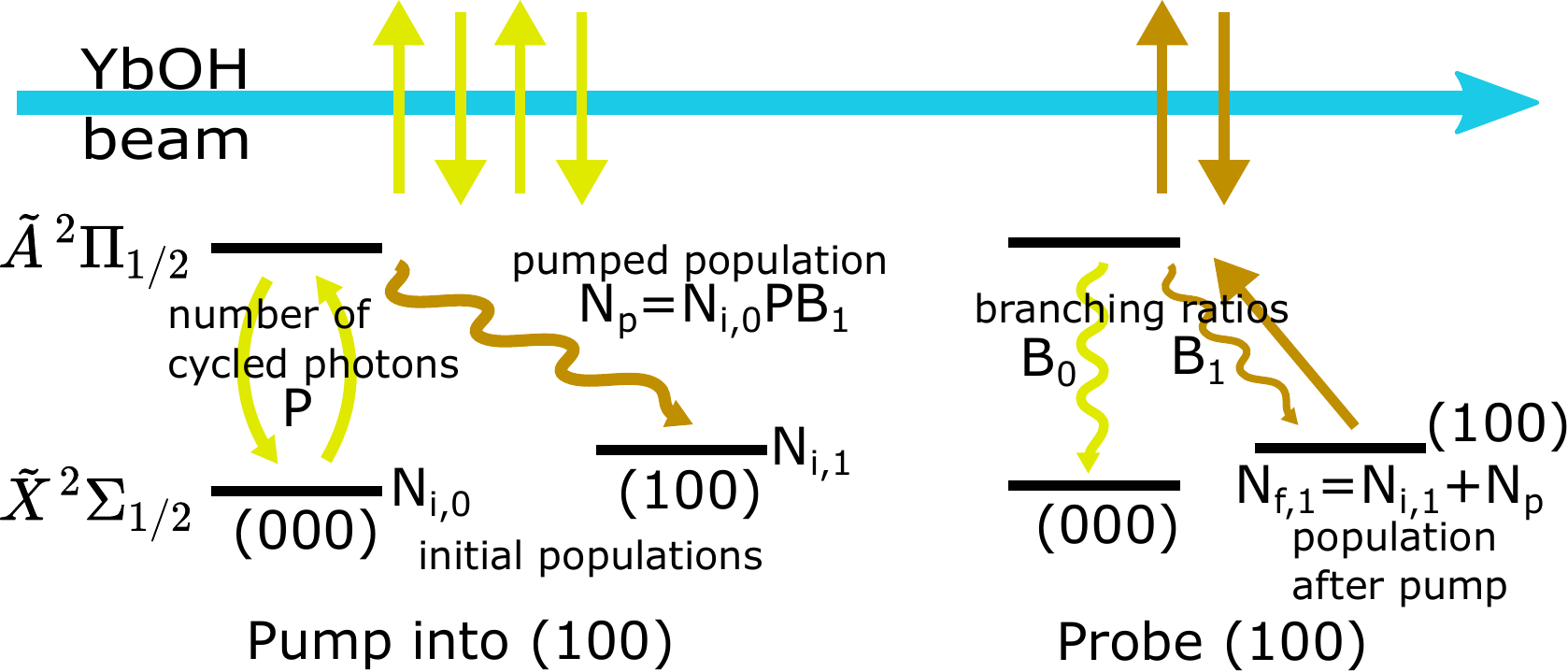}
\caption{Schematic for measuring the number of photon cycled by measuring the population transfer into one of the hyperfine levels of the first excited vibration state (1,0,0), or alternatively labeled $v\,{=}\,1$. }
\label{ppfig}
\end{figure*}

To verify that we have achieved rotational closure, we optically cycle and measure the number of photons scattered per molecule. If we have addressed all branching within $X$, the molecules will cycle until they are optically pumped into a dark vibrational state. The majority of molecules are pumped into one of the hyperfine levels of the $\Tilde{X}^2\Sigma^+(1,0,0)$ state with one quantum of Yb-O stretching motion ($v\,{=}\,1$), which has a branching ratio of $9.11(55)\%$ from $A$. After cycling upstream in the molecular beam, we measure the population downstream in one of the hyperfine levels of the $v''\,{=}\,1$ state (see fig.~\ref{ppfig}). 

The main benefit of such a population transfer measurement, compared to a direct fluorescence measurement of optical cycling, is the ability to reject accidental fluorescence from nearby transitions. There are 5 isotopes of Yb with significant abundance, and both $\prescript{171,173}{}{}$YbOH have extra complexity arising from their nuclear spins. As a result, the band head region where the cycling transitions are located is extremely congested with additional lines, and a direct fluorescence measurement would make it difficult to disentangle the effects of photon cycling versus merely addressing multiple lines in different states and isotopologues.

By using a pump-probe scheme to measure the population transfer, most of these unwanted transitions do not contribute to our results. The exception is when both pump and probe coincide with optical pumping of a different isotopologue or level that is unrelated to cycling. By using both resonant chemical enhancement \cite{Jadbabaie2020} of a specific Yb isotope and intentional selection of the hyperfine level in $v''\,{=}\,1$ for the probe transition, contaminant contributions to optical pumping are relatively small, and can be estimated using detunings and branching ratios. 

Our figure of merit is the quantity $D$, defined as the difference in $v''\,{=}\,1$ probe fluorescence caused by the cycling pump laser, normalized against probe fluorescence without the pump. Because the $v''\,{=}\,1 \rightarrow v'\,{=}\,0$ probe transition has very small branching ratio of $0.0911$, we operate with low saturation parameter, and the measurement of fluorescence increase is equivalent to an increase in population. To extract the number of photons cycled from $D$, we use the following relationship:

\begin{equation}
\begin{split}
D & \,{=}\,\frac{\text{fluorescence}_{\text{pump}}-\text{fluorescence}_{\text{no pump}}}{\text{fluorescence}_{\text{no pump}}} \\
&\approx\frac{N_{f,1}-N_{i,1}}{N_{i,1}} \,{=}\, \frac{N_P}{N_{i,1}}\approx \frac{N_{i,0}\cdot P \cdot B_1}{N_{i,1}},
\end{split} 
\end{equation}
where $N_P$ is the population transferred into $v''\,{=}\,1$ from optical pumping, $N_{i,1}$ is the initial population of the $v''\,{=}\,1$ state being probed, $N_{i,0}$ is the initial population of the $v''\,{=}\,0$ states being pumped, $P\approx1/(1-B_0)$ is the number of photons cycled per pumped molecule, $B_0$ is the sum of total branching ratios of all the pumped states in $v''\,{=}\,0$, and $B_1$ is the branching ratio down to the specific hyperfine state probed in $v''\,{=}\,1$ probe state (see fig.~\ref{ppfig}.)

The ratio $N_{i,0}/N_{i,1}$ can be calculated by making two assumptions about the initial populations of the relevant states: first, that the states within a vibrational manifold are well thermalized coming out of the cryogenic buffer gas beam source, and second, that the ratios of population between vibration ground and first excited vibration mode, $R\,{=}\,N_{total(v\,{=}\,1)}/N_{total(v\,{=}\,0)}$, is the same for different isotopologues of YbOH under the same source condition. These assumptions are supported by tests done with $\prescript{174}{}{}$YbOH and and from behavior in other such molecular sources \cite{Hutzler2012CBGB}.

\subsection{Calibration with $\prescript{174}{}{}$YbOH}

\begin{figure}[t]
    \centering
    \includegraphics[width=0.8\columnwidth]{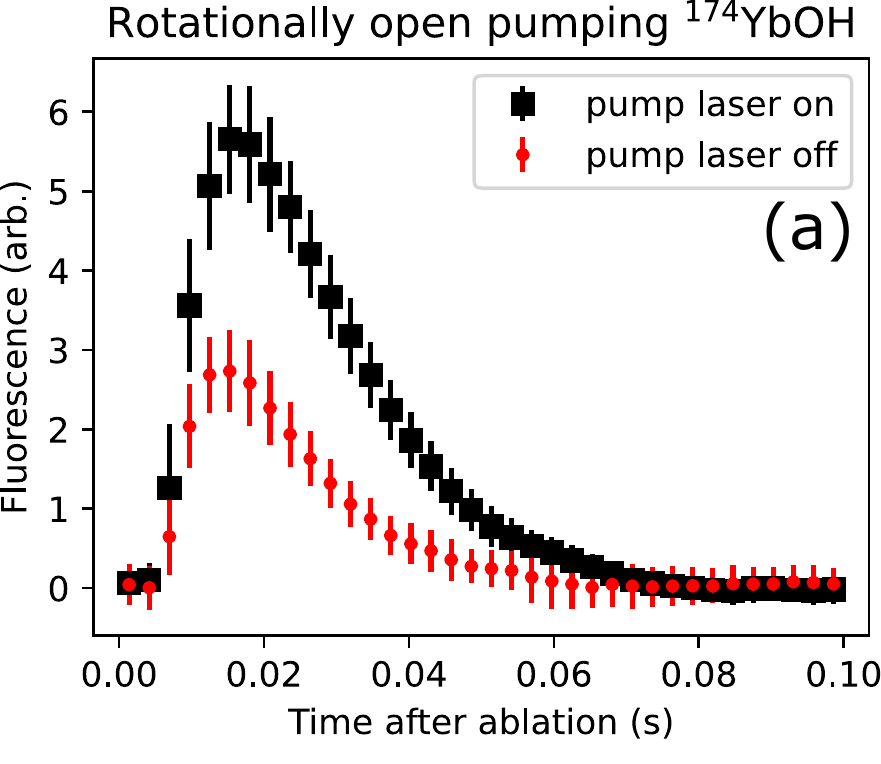}
    \includegraphics[width=0.8\columnwidth]{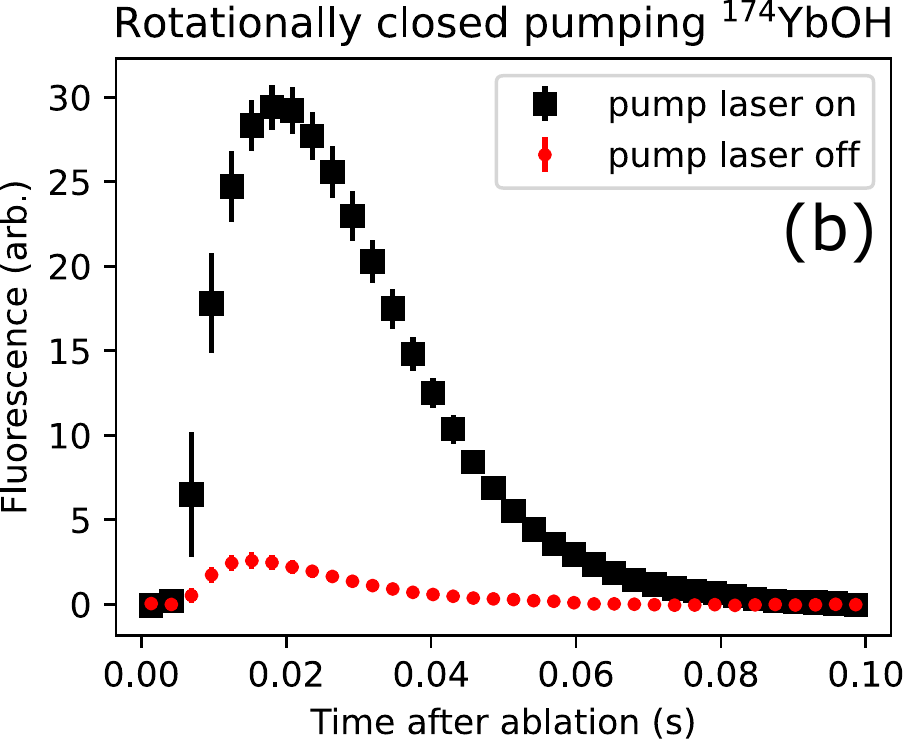}
    \caption{$\prescript{174}{}{}$YbOH fluorescence signals from a probe laser addressing the $v''\,{=}\,1$ $N''\,{=}\,1$ $J''\,{=}\,1/2$ state. Error bars represent 1-$\sigma$ variation due shot-to-shot fluctuations over 80 beam pulses. (a) The pump laser has no sidebands and only addresses the $J''\,{=}\,1/2$ state in the ground spin-rotation pair. The increase in fluorescence is a factor of $D_{ro}\,{=}\,1.38\pm 0.23$, as described in the main text. (b) The pump laser addresses both $J''\,{=}\,1/2$ and $J''\,{=}\,3/2$, achieving rotationally closed cycling and increasing the fluorescence by $D_{rc}\,{=}\,15.0\pm 1.6$.  }
    \label{r174fig}
\end{figure}

We first made measurements with $\prescript{174}{}{}$YbOH to validate our method, since photon cycling in $\prescript{174}{}{}$YbOH has been carefully characterized elsewhere~\cite{Zhang2021BR}, as well as to derive the population ratio $R$ between the vibrational modes in our molecular beam. Results are shown in fig.~\ref{r174fig}, which gives two final $D$ values for two different pumping laser configurations: $D_{ro}$ for rotationally open (ro) pumping, and $D_{rc}$ for rotationally closed (rc) pumping. For $\prescript{174}{}{}$YbOH, we only need one sideband to cover the spin-rotation splitting and achieve rotationally closed cycling \cite{Augenbraun2020YbOHLC}, which gives a scattered photon number of about $P\approx1/(1-B_0)\,{=}\,9.1$ per addressed molecule, limited only by the vibrational branching ratio $B_0\,{=}\,0.89$. In comparison, when only addressing the $J\,{=}\,1/2$ state of the spin-rotation pair, which has a rotational branching of $0.67$, the photon number expected is only $P\,{=}\,2.5$, where $B_0\,{=}\,0.89\times0.67\,{=}\,0.60$.

With the assumption of thermalization, we can treat all rotational levels and $M_F$ sublevels within a given vibronic manifold as having equal initial population since the level splittings are all far smaller than the thermal energy $k_B T\approx30$~GHz, where $k_B$ is Boltzmann's constant and our operating temperature is $T\,{\approx}\,1.5$ K. Let the initial population of the pumped state be 1 times their degeneracies (including hydrogen nuclear spin), which, for $\prescript{174}{}{}$YbOH, means $N_{i,0}\,{=}\,12$ for addressing both of the spin-rotation (SR) pair, and $N_{i,0}\,{=}\,4$ for addressing only the $J''\,{=}\,1/2$ state of the pair. Then, using $R\,{=}\,N_{total(v\,{=}\,1)}/N_{total(v\,{=}\,0)}$ as an unknown variable, we have the initial population of the probed state as $N_{i,1}\,{=}\,4R$. Here the probe laser addresses the $v''\,{=}\,1 ,N''\,{=}\,1 ,J''\,{=}\,1/2$ state, which has a total branching ratio of $B_1\,{=}\,0.06$ coming from $J'\,{=}\,1/2(+)$ excited state. 

As a result, when the pump addresses both components of the SR pair, we would expect the measurement to be $D_{rc}\,{=}\,N_{i,0} P  B_1/N_{i,1}\,{=}\,1.64/R$.  Similarly, when the pump addresses only the $J''\,{=}\,1/2$ state, we expect $D_{ro}\,{=}\,0.15/R$ . Experimentally, we measured $D$ to be $D_{rc}\,{=}\,15.0\pm 1.6$, and $D_{ro}\,{=}\,1.38\pm 0.23$, which gives $R$ values of $0.11\pm 0.01$ and $0.11\pm 0.02$. In other words, the $v''\,{=}\,1$ manifold has approximately $11\%$ the population of the $v''\,{=}0$ manifold. The results are averaged from 40 data sets, where the statistical uncertainties mainly come from shot to shot variations of the molecule source. The two different pumping configurations give the same $R$ value within our error, validating the measurement method. For the measurements on $\prescript{171,173}{}{}$YbOH, we will similarly perform both rotationally closed and open pumping, but will use the $R\,{=}\,0.11$ as known parameter to derive the number of scattered photons per molecule.

\subsection{Spectroscopy of $\prescript{171,173}{}{}$YbOH $\Tilde{X}^2\Sigma^+(1,0,0)$}

\begin{figure}[t]
    \centering
    \includegraphics[width=0.9\columnwidth]{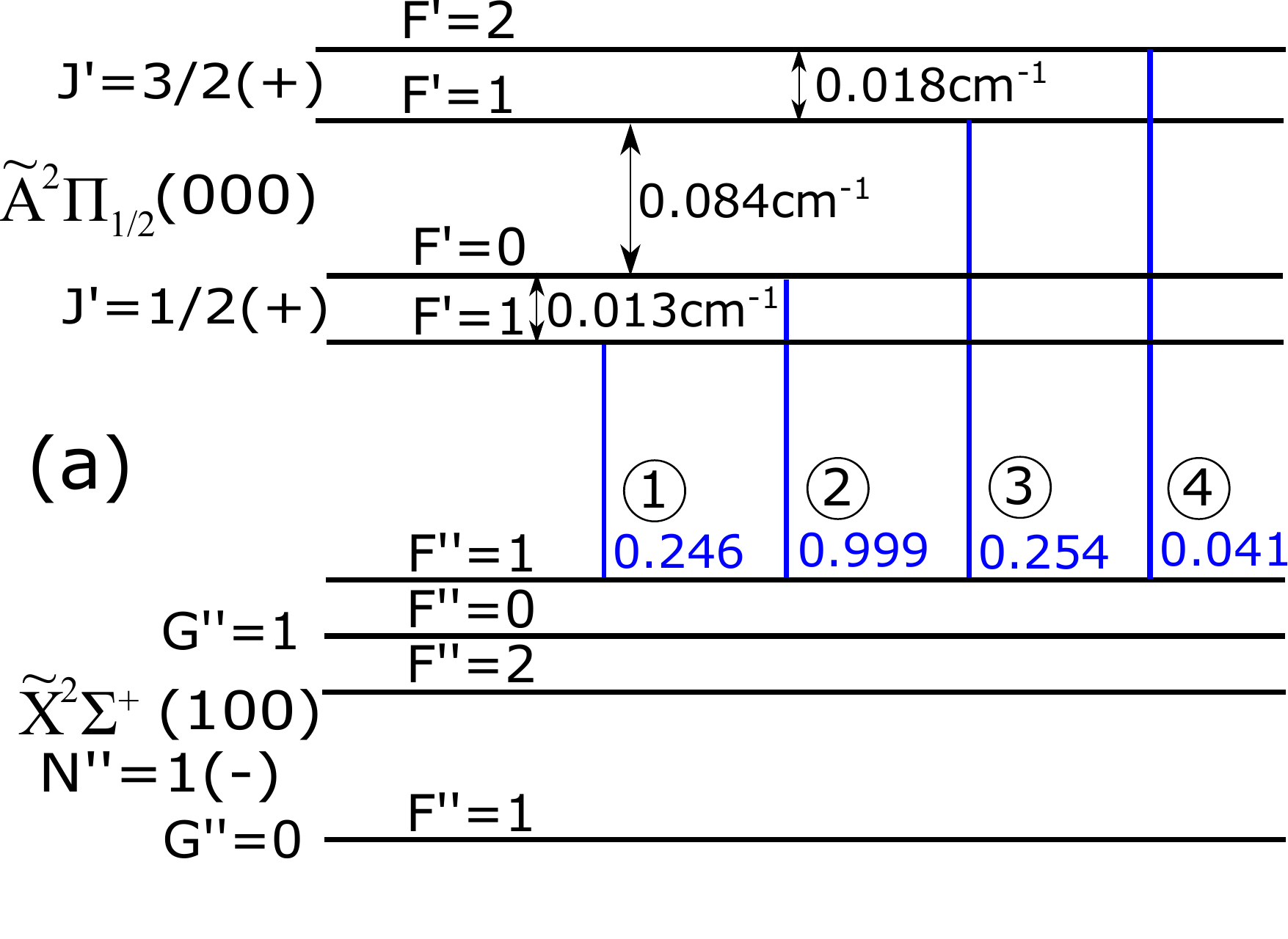}
    \includegraphics[width=\columnwidth]{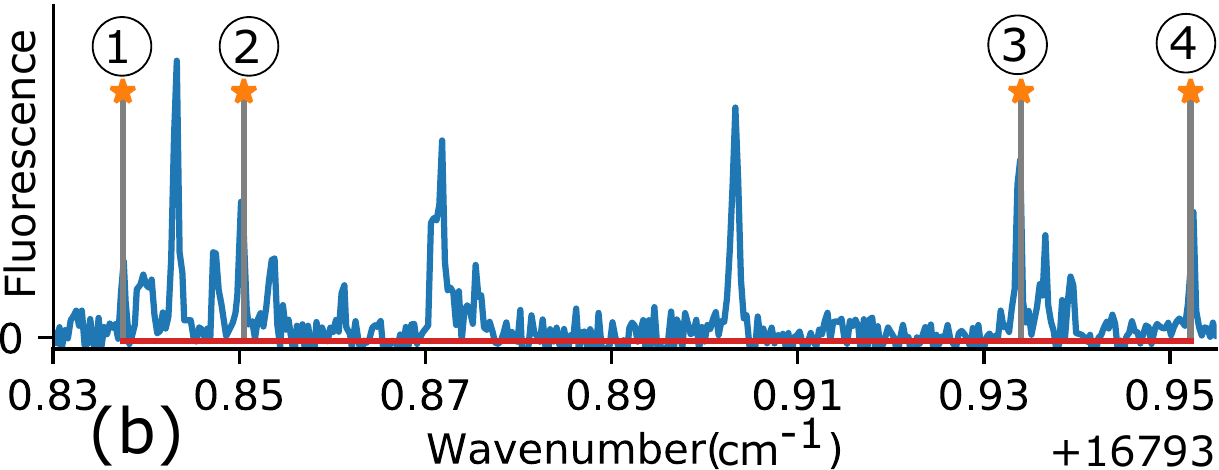}
    \caption{Example of $v''=1$ spectroscopy for $\prescript{171}{}{}$YbOH. Transitions are identified by matching spectral splittings to known excited state splittings~\cite{Pilgram2021YbOHOdd}. (a) is the level diagram of the involved ground and excited states, with calculated branching ratios for the target transitions. (b) is the measured fluorescence spectrum with the same four transitions marked in (a). Some of the unmarked peaks nearby are transitions from $F''\,{=}\,0,2$ ground hyperfine levels to the same excited states, identifiable via the common splittings.}
    \label{s171fig}
\end{figure}

In order to measure the population transfer from optical pumping, it is important to establish the correct quantum numbers and energies for the  $\Tilde{X}^2\Sigma^+(1,0,0)$, or $v''\,{=}\,1$, states within the $N''\,{=}\,1$ manifold for $\prescript{171,173}{}{}$YbOH. Given the known odd and even isotopologue parameters for the $v''\,{=}\,0$ state~\cite{Pilgram2021YbOHOdd,nakhate2019pure}, and the even isotopologue parameters for the $v''\,{=}\,1$ state~\cite{steimle2019field}, we first predict the expected spectral region for the $\Tilde{X}^2\Sigma^+(1,0,0)\rightarrow \Tilde{A}^2\Pi_{1/2}(0,0,0)$ transition. We then measure spectra around the predicted region and look for splittings that match the known excited splittings between hyperfine levels of $\Tilde{A}^2\Pi_{1/2}(0,0,0)$ $J\,{=}\,1/2(+)$ and $J\,{=}\,3/2(+)$. For transitions with a shared ground state and different excited state, the observed line splittings will match the known structure of the excited $\Tilde{A}^2\Pi_{1/2}(0,0,0)$ state \cite{Pilgram2021YbOHOdd}.

The spectral region is congested with overlapping transitions from multiple rotational and hyperfine states, as well as from other isotopologues. Furthermore, the $v\,{=}\,1$ states have an order of magnitude smaller population coming out of the cryogenic buffer gas source compared to $v\,{=}\,0$, as we learned from the $\prescript{174}{}{}$YbOH measurements in the previous section. In order to differentiate the multiple transitions, we used chemical enhancement \cite{Jadbabaie2020} and population pumping from $v\,{=}\,0$.

As an example, fig.~\ref{s171fig} shows four transitions from the $X, v''=1, N''\,{=}\,1 , G''\,{=}\,1 , F''\,{=}\,1$ ground state to four hyperfine levels in the excited $A, v'=0, J'\,{=}\,1/2$ and $J'\,{=}\,3/2$ states. The spectrum in fig.~\ref{s171fig}(b) is taken downstream of a power-broadened pumping laser with EOM-generated sidebands that roughly addresses all the $v''\,{=}\,0 , N''\,{=}\,1$ ground states connected to the $J'\,{=}\,1/2(+) , F'\,{=}\,1$ excited state. The EOM sideband scheme is not optimized here because we need initial spectroscopy to confirm our probe state selection. After the prerequisite spectroscopy, we are able to optimize the sideband scheme to achieve rotationally closed photon cycling.

A full spectroscopic analysis and parameter fit for all the $v\,{=}\,1$ states is beyond the scope of this work. However, we were able to identify all of the relevant  hyperfine levels within the $v''\,{=}\,1, \, N''\,{=}\,1$ manifold for $\prescript{171,173}{}{}$YbOH (see supplement). We are confident in our state identification based on the matching splittings in both ground and excited states, and based on the observation of expected population transfers.

\subsection{Photon cycling in $\prescript{171,173}{}{}$YbOH}

After identifying all of the $v''\,{=}\,1$, $N''\,{=}\,1$ states for both $\prescript{171,173}{}{}$YbOH, we decided to use the $\Tilde{X}^2\Sigma^+(1,0,0) \ket{N''\,{=}\,1 , G''\,{=}\,1 , F''\,{=}\,1} \rightarrow \Tilde{A}^2\Pi_{1/2}(0,0,0)\ket{J'\,{=}\,1/2(+) , F'\,{=}\,0}$ transition for the $\prescript{171}{}{}$YbOH probe, since it is the rotationally-closed analogue of the corresponding $(0,0,0)\rightarrow(0,0,0)$ transition, and relatively well separated from other hyperfine levels. For $\prescript{173}{}{}$YbOH, we decided to use the $\Tilde{X}^2\Sigma^+(1,0,0)\ket{1,3,2} \rightarrow\Tilde{A}^2\Pi_{1/2}(0,0,0)\ket{1/2(+), 2}$ transition, as it minimized accidental pump/probe signal contaminants from unwanted states, and since it has very similar branching from both excited $F'$ hyperfine levels, simplifying data analysis and modeling. With the probe transitions identified, we were able to optimize the cycling schemes, and lower the pump laser power required to less than 25~mW per sideband in a 5~mm diameter beam.

\begin{figure}[t]
    \centering
    \includegraphics[width=0.9\columnwidth]{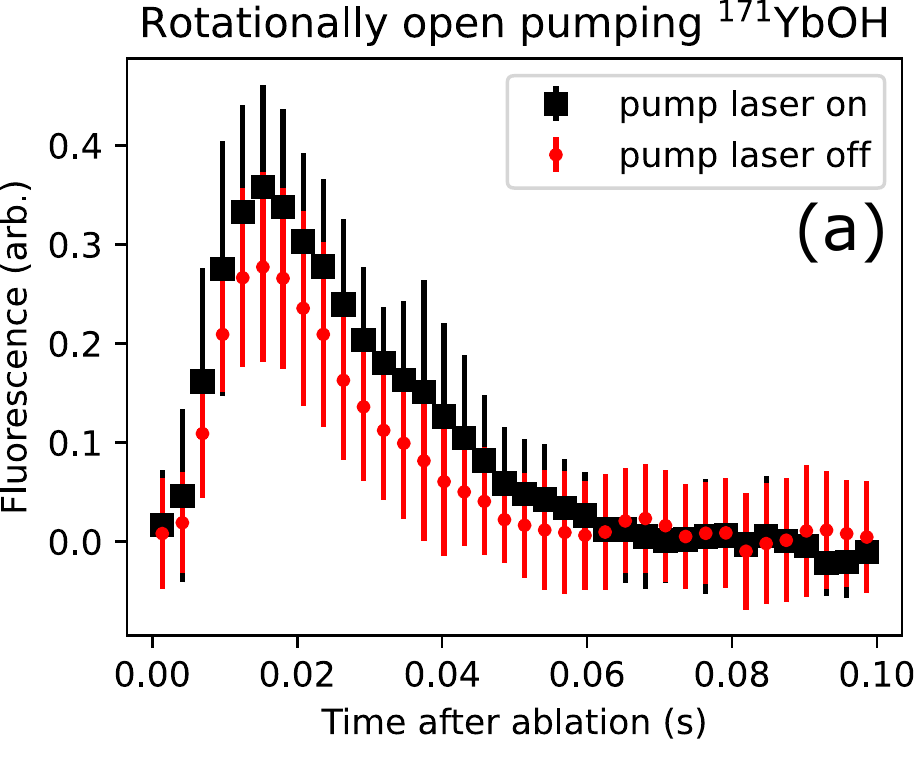}
    \includegraphics[width=0.9\columnwidth]{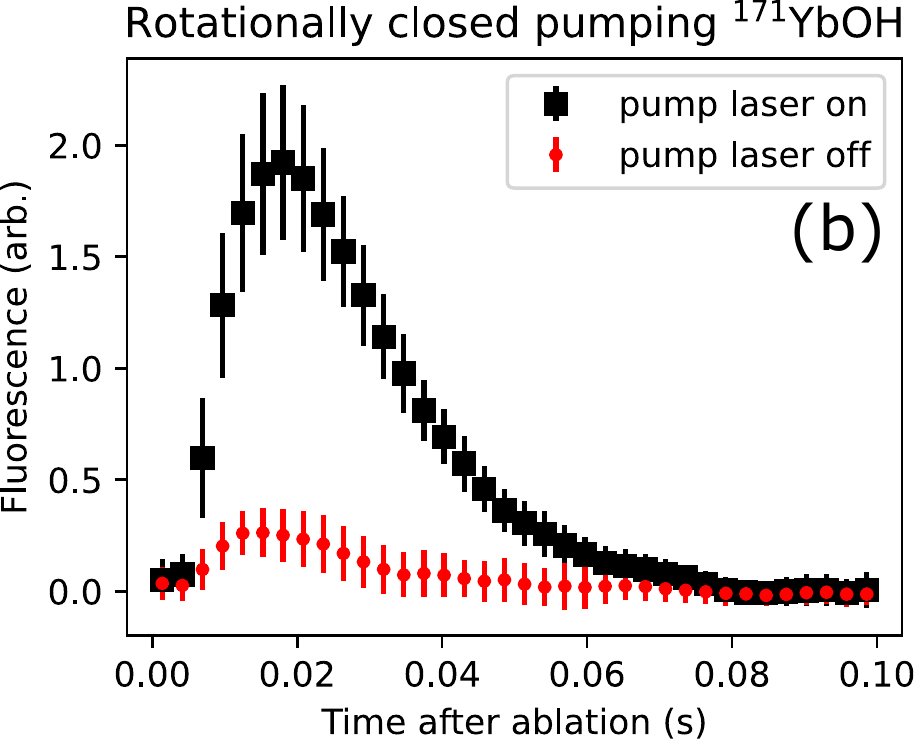}
    \caption{$\prescript{171}{}{}$YbOH fluorescence signals from the probe laser addressing $v''\,{=}\,1$ $\ket{N''\,{=}\,1 , G''\,{=}\,1 , F''\,{=}\,1}$. Error bars represent $1\sigma$ spreads due shot-to-shot fluctuations over 80 beam pulses. (a) Pump laser addressing the $F''\,{=}\,1 \rightarrow F'\,{=}\,1$ rotationally open transition. The increase in fluorescence is $D_{ro}\,{=}\,0.48\pm 0.17$. (b) Pump laser addressing the $F''\,{=}\,1 \rightarrow F'\,{=}\,0$ rotationally closed cycling transition. The fluorescence increase is  $D_{rc}\,{=}\,7.4\pm 1.3$.  }
    \label{r171fig}
\end{figure}

\begin{figure}[t]
    \centering
    \includegraphics[width=0.9\columnwidth]{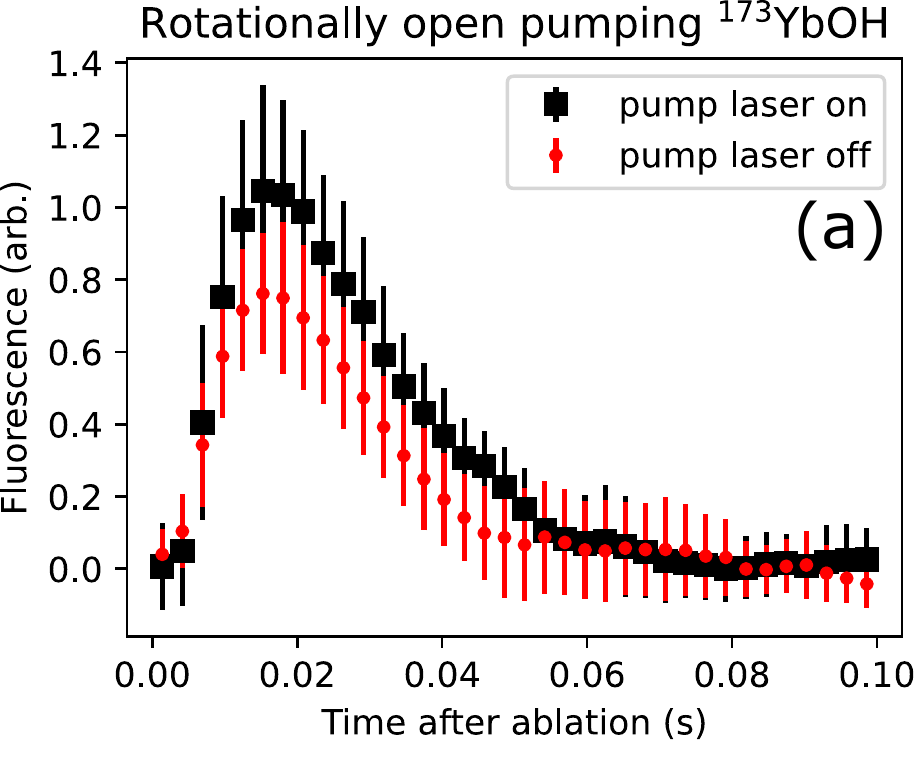}
    \includegraphics[width=0.9\columnwidth]{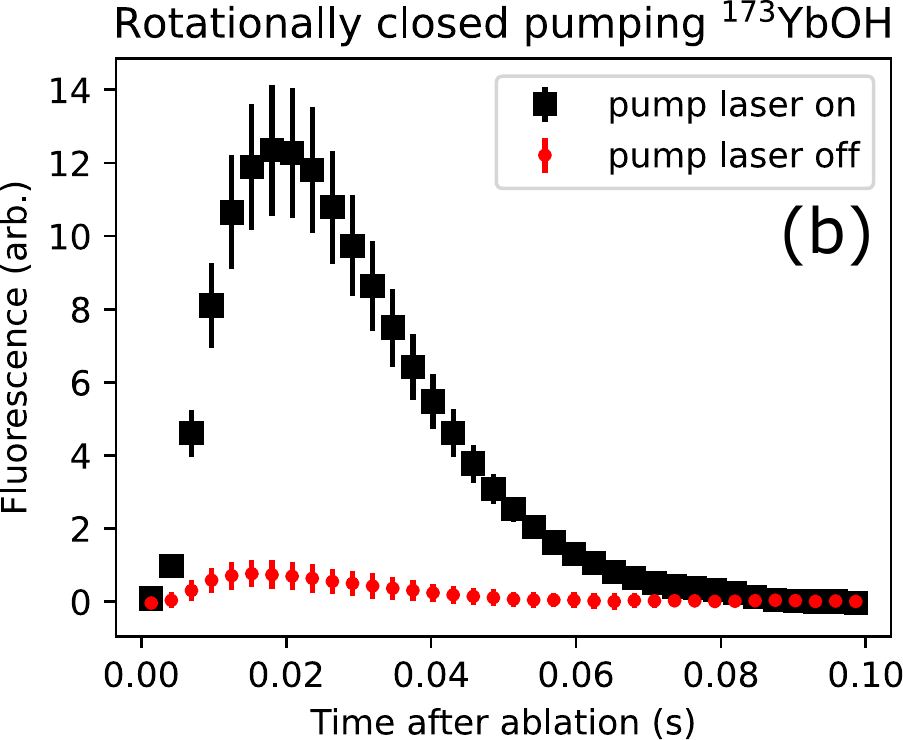}
    \caption{$\prescript{173}{}{}$YbOH fluorescence signals from  probe laser addressing $v''\,{=}\,1$ $\ket{N''\,{=}\,1 , G''\,{=}\,3 , F''\,{=}\,2}$. Error bars represent $1\sigma$ spreads due shot-to-shot fluctuations over 80 beam pulses. (a) Pump laser with no sidebands, addressing only the $\ket{1,2,3}\rightarrow F'\,{=}\,3$ transition. The change in integrated fluorescence is $D_{ro}\,{=}\,0.40\pm 0.14$. (b) The pump laser addresses all the $v''\,{=}\,0, N''\,{=}\,1$ hyperfine levels using the optimized EOM scheme, and thus achieves rotationally closed cycling. The increase in fluorescence is $D_{rc}\,{=}\,13.0\pm 1.9$.  }
    \label{r173fig}
\end{figure}

Following the same procedure used for $\prescript{174}{}{}$YbOH laid out in the previous section, we measured the scattered photon number per molecule for $\prescript{171}{}{}$YbOH and $\prescript{173}{}{}$YbOH to confirm that our cycling schemes work as expected. Fig.~\ref{r171fig} shows the results for $\prescript{171}{}{}$YbOH. The rotationally closed cycling transition is from $F''\,{=}\,1$ to $F'\,{=}\,0$, which gives a initial population, including unresolved -OH hyperfine degeneracy, of $N_{i,0}\,{=}\,6$. The pump transition for the rotationally open comparison is from the same ground state, hence same initial population, but different excited state, $F'\,{=}\,1$, which gives a calculated branching ratio of $B_0\,{=}\,0.22$. The probe transition addresses the equivalent ground state in $v''\,{=}\,1$, so $N_{i,1}\,{=}\,6R$, and the total branching ratio from $A$ to $v''\,{=}\,1$ is $B_1\,{=}\,0.09$ from $F'\,{=}\,0$ and $B_1\,{=}\,0.02$ from $F'\,{=}\,1$. 

Fig.~\ref{r171fig}(a) shows results from pumping on the rotationally open transition, which gives an increase in fluorescence of $D_{ro}\,{=}\,0.48\pm 0.17$. Using the value $R\,{=}\,0.11\pm 0.01$ from $\prescript{174}{}{}$YbOH, we obtain a photon number of $2.4\pm 0.8$, which is larger than the $1.3$ expected from a branching ratio of $0.22$. The reason for the discrepancy is contribution from nearby $v''\,{=}\,1$ ground state hyperfine levels which are 80 and 150~MHz away. The nearby levels contribute both to the initial population $N_{i,0}$ and to the branching ratio $B_0$. By calculating and including the effect of photon scattering from these nearby states, we actually expect a fluorescence increase of $D_{ro}\,{=}\,0.45$, which fits the data well and thus confirms that the assumption holds that the vibrational population ratio $R$ is isotope independent. For the rotationally closed transition, we measured a fluorescence increase $D_{rc}\,{=}\,7.4\pm 1.3$, which gives a scattered photon number of $8.9\pm 1.5$, which matches the expectation of $9.1$ from the vibrational branching ratio of $0.89$.

Fig.~\ref{r173fig} shows results for $\prescript{173}{}{}$YbOH. The reference pump transition addressing $F''\,{=}\,3$ has $N_{i,0}\,{=}\,14$, and a branching ratio of $B_1\,{=}\,0.02$ to the probe state in $v''\,{=}\,1$. The cycling pump addresses all 6 ground hyperfine levels, giving $N_{i,0}\,{=}\,72$, and the same branching ratio to the probe state. The probe laser addresses $F''\,{=}\,2$, giving $N_{i,1}\,{=}\,10R$. With the same $R=0.11$ value as before, the measured normalized difference in fluorescence, $D_{ro}\,{=}\,0.40\pm 0.14$, means that the reference pump scattered $P\,{=}\,1.4\pm 0.5$ photons per molecules, matching the expectation of $1.5$ for the branching ratio of $B_0\,{=}\,0.33$ on the open transition. For rotationally closed cycling, the result of $D_{rc}\,{=}\,13.0\pm 1.9$ would indicate a photon number of $9.1\pm 1.3$, matching the expected $9.1$ from purely vibrational branching.  Note that there is an expected rotational leakage of 0.5\% due to hyperfine-induced transitions~\cite{Pilgram2021YbOHOdd}, which is below our resolution to observe; however, this could be addressed by adding additional sidebands to address the relevant $N''=3$ levels.

Table \ref{P_table} summarizes the result of our photon cycling measurements. Our measured results for the number of photons scattered per molecule, $P$, are in good agreement with the expectation from our calculated branching ratios. In particular, once rotational branching is addressed, the value of $P$ is limited by vibrational branching, which is independent of isotopologue. This is also reflected in our data, with each isotopologue having the same value of $P$ for rotationally closed pumping. 

\begin{table}[h]
\centering
\begin{tabular}{lccl} 
\hline\hline
Molecule & RO/RC  & Expected $P$ & Measured $P$ \\ 
\hline
\multirow{2}{*}{$^{174}$YbOH} & RO  & 2.5 &  \\
 & RC  & 9.1 &  \\ 
\hline
\multirow{2}{*}{$^{171}$YbOH} & RO  & 1.3 & 2.4 ± 0.8 \\
 & RC & 9.1 & 8.9 ± 1.5 \\ 
\hline
\multirow{2}{*}{$^{173}$YbOH} & RO   & 1.5 & 1.4 ± 0.5 \\
 & RC  & 9.1 & 9.1 ± 1.3 \\
\hline\hline
\end{tabular}
\caption{A comparison of expected and measured numbers of photons scattered per molecule, denoted as $P$. Here, RO refers to rotationally open pumping, and RC refers to rotationally closed pumping. Measurements of $P$ in $^{174}$YbOH are used, along with the theoretical prediction, to derive the vibration population ratio $R$. Both RO and RC for $^{174}$YbOH yielded the same resulting $R$. The high $P$ measured in RO $^{171}$YbOH is caused by nearby ground hyperfine levels contributing.}
\label{P_table}
\end{table}

\subsection{Conclusion}
We have demonstrated a scheme that achieves rotationally closed photon cycling of $\prescript{171}{}{}$YbOH and $\prescript{173}{}{}$YbOH, both of which have complicated hyperfine structure. The method is technically straightforward and can be easily extended to higher-order vibrational repumps to enable larger and larger numbers of scattered photons.  Since molecule-based studies of the nuclear weak force and the search for CP-violating nuclear magnetic quadrupole moments require species with significant electron density at a heavy, spinful nucleus, the ability to address the resulting complicated hyperfine structure will be important for future experimental searches.

\begin{acknowledgments}

We acknowledge helpful discussions with Nickolas Pilgram.  This work was supported by the Gordon and Betty Moore Foundation (GBMF7947), the Alfred P. Sloan Foundation (G-2019-12502), the Heising-Simons Foundation (2019-1193 and 2022-3361), and an NSF CAREER award (PHY-1847550).  P. Y. acknowledges support from the Eddleman Graduate Fellowship through the Institute for Quantum Information and Matter (IQIM).

\end{acknowledgments}

\bibliography{references}

\newpage

\hspace{1in}

\newpage

\subsection{Appendix A: tables of measured lines}

\begin{table}[H]
\begin{tabular}{|c|c|c|c|c|} \hline
\;$G''$ \; & \;  $F''$ \; & \;$J'(P)$\; & \;$F'$\; & Wavenumber/cm$^{-1}$ \\ \hline
1   & 1   & 0.5+  & 1  & 16793.8375       \\ 
1   & 0   & 0.5+  & 1  & 16793.8400       \\ 
1   & 2   & 0.5+  & 1  & 16793.8433       \\ 
1   & 1   & 0.5+  & 0  & 16793.8505       \\ 
1   & 0   & 0.5+  & 0  & 16793.8530       \\ 
1   & 2   & 0.5+  & 0  & 16793.8563       \\ 
1   & 1   & 1.5+  & 1  & 16793.9340       \\ 
1   & 0   & 1.5+  & 1  & 16793.9365       \\ 
1   & 2   & 1.5+  & 1  & 16793.9398       \\ 
1   & 1   & 1.5+  & 2  & 16793.9522       \\ 
1   & 0   & 1.5+  & 2  & 16793.9547       \\ 
1   & 2   & 1.5+  & 2  & 16793.9580       \\ 
0   & 1   & 0.5+  & 1  & 16794.0698       \\ 
0   & 1   & 0.5+  & 0  & 16794.0828       \\ 
0   & 1   & 1.5+  & 1  & 16794.1663       \\ 
0   & 1   & 1.5+  & 2  & 16794.1845       \\ \hline
\end{tabular}
\caption{Measured transitions between $X''(100) N''=1$ and $A(000)$ in $^{171}$YbOH.  Uncertainties are estimated to be 0.0005~cm$^{-1}$ due to wavemeter drift and uncertainty}
\end{table}

\begin{table}[H]
\begin{tabular}{|c|c|c|c|c|} \hline
\;$G''$ \; & \;  $F''$ \; & \;$J'(P)$\; & \;$F'$\; & Wavenumber/cm$^{-1}$ \\ \hline
2 & 1 & 0.5+ & 2 & 16794.0030 \\
2 & 3 & 0.5+ & 2 & 16794.0100 \\
2 & 3 & 0.5+ & 3 & 16794.0151 \\
2 & 2 & 0.5+ & 2 & 16794.0312 \\
2 & 2 & 0.5+ & 3 & 16794.0363 \\
2 & 3 & 1.5+ & 4 & 16794.1098 \\
2 & 1 & 1.5+ & 2 & 16794.1183 \\
2 & 2 & 1.5+ & 3 & 16794.1353 \\
2 & 2 & 1.5+ & 1 & 16794.1623 \\
3 & 2 & 0.5+ & 2 & 16794.1891 \\
3 & 2 & 0.5+ & 3 & 16794.1942 \\
3 & 4 & 0.5+ & 3 & 16794.2068 \\
3 & 3 & 0.5+ & 2 & 16794.2230 \\
3 & 3 & 0.5+ & 3 & 16794.2281 \\
3 & 4 & 1.5+ & 4 & 16794.3010 \\
3 & 2 & 1.5+ & 2 & 16794.3043 \\
3 & 4 & 1.5+ & 3 & 16794.3061 \\
3 & 2 & 1.5+ & 1 & 16794.3204 \\
3 & 3 & 1.5+ & 3 & 16794.3274 \\
3 & 3 & 1.5+ & 2 & 16794.3383 \\\hline
\end{tabular}
\caption{Measured transitions between $X''(100) N''=1$ and $A(000)$ in $^{173}$YbOH.   Uncertainties are estimated to be 0.0005~cm$^{-1}$ due to wavemeter drift and uncertainty}
\end{table}

\newpage
\subsection{Appendix B: tables of Calculated $X(100)$ energy levels}

\begin{table}[H]
\begin{tabular}{|c|c|c|c|} \hline
 \;\; $N$ \;\; &  \;\; $G$ \;\; & \;\;  $F$ \;\; & Wavenumber/cm$^{-1}$ \\ \hline
1 & 0 & 1 & 529.2047  \\
1 & 1 & 2 & 529.4370  \\
1 & 1 & 0 & 529.4345  \\
1 & 1 & 1 & 529.4312  \\\hline
\end{tabular}
\caption{Energies of levels in $^{171}$YbOH $X(100)$ determined by this work.    Uncertainties are estimated to be 0.0005~cm$^{-1}$ due to wavemeter drift and uncertainty, and excited state energy uncertainty.}
\end{table}

\begin{table}[H]
\begin{tabular}{|c|c|c|c|} \hline
 \;\; $N$ \;\; &  \;\; $G$ \;\; &  \;\; $F$ \;\; & Wavenumber/cm$^{-1}$ \\ \hline
1 & 2 & 1 & 530.146  \\
1 & 2 & 3 & 530.139  \\
1 & 2 & 2 & 530.118  \\
1 & 3 & 2 & 529.960  \\
1 & 3 & 4 & 529.947  \\
1 & 3 & 3 & 529.926  \\\hline
\end{tabular}
\caption{Energies of levels in $^{173}$YbOH $X(100)$ determined by this work.   Uncertainties are estimated to be 0.001~cm$^{-1}$ due to wavemeter drift and uncertainty, and excited state energy uncertainty.}
\end{table}

\end{document}